\documentclass[11pt,a4paper]{article}

\usepackage{graphicx}
\usepackage{latexsym}
\usepackage{amsmath,amsfonts,amsthm}
\newcommand{\vecr}{\mathbf{r}}

\begin{document}
\pagenumbering{roman} \setcounter{page}{0} 
\title{Scaling of Self-Avoiding Walks in High Dimensions}
\author{A. L. Owczarek\ddag  \, and T. Prellberg\dag\thanks{{\tt {\rm email:}
aleks@ms.unimelb.edu.au,thomas.prellberg@tu-clausthal.de}} \\
        \ddag Department of Mathematics and Statistics,\\
         The University of Melbourne,\\
         3010, Australia.\\
        \dag Institut f\"ur Theoretische Physik,\\
        Technische Universit\"at Clausthal,\\ Arnold Sommerfeld Stra\ss e 6,\\ D-38678 Clausthal-Zellerfeld,\\ Germany
 }
\date{April 9, 2001
}

\maketitle 
 
\begin{abstract} 

We examine self-avoiding walks in dimensions 4 to 8 using
high-precision Monte-Carlo simulations up to length $N=16384$,
providing the first such results in dimensions $d > 4$ on which we
concentrate our analysis.  We analyse the scaling behaviour of the
partition function and the statistics of nearest-neighbour contacts,
as well as the average geometric size of the walks, and compare our
results to $1/d$-expansions and to excellent rigorous bounds that
exist. In particular, we obtain precise values for the connective
constants, $\mu_5=8.838544(3)$, $\mu_6=10.878094(4)$,
$\mu_7=12.902817(3)$, $\mu_8=14.919257(2)$ and give a revised estimate
of $\mu_4=6.774043(5)$.  All of these are by at least one order of
magnitude more accurate than those previously given (from other
approaches in $d>4$ and all approaches in $d=4$).  Our results are
consistent with most theoretical predictions, though in $d=5$ we find
clear evidence of anomalous $N^{-1/2}$-corrections for the scaling of
the geometric size of the walks, which we understand as a non-analytic
correction to scaling of the general form $N^{(4-d)/2}$ (not present
in pure Gaussian random walks).

\vspace{1cm} 
 

\end{abstract} 

\vfill

\newpage

\pagenumbering{arabic}
\setcounter{page}{1}

\section{Introduction}

The universal properties of linear flexible polymers in a dilute
solution can be modelled by the lattice model of self-avoiding walks
(SAW), and have been studied for this purpose \cite{cloizeaux1990a-a}
for over 50 years \cite{orr1946a-a}. Being described by the limit
$n\rightarrow 0$ of the O($n$) $\phi^4$ field theory
\cite{gennes1972a-a,gennes1979a-a} places SAW amongst the most
fundamental models in statistical physics along with the Ising ($n=1$)
and Heisenberg ($n=3$) models, while its description as a non-Markovian
random walk makes it of intense interest to mathematicians
\cite{madras1993a-a}. SAW are also of interest to the combinatorial
mathematician as a fundamental combinatorial problem. As a critical
phenomenon, in the context of theoretical physics, the limit of the
length of the walk going to $\infty$ can be thought of as equivalent
to approaching a critical temperature (in a generating function
approach the generating variable acts as the Boltzmann weight in an
O($n$) model). Results from the associated field theory
\cite{gennes1972a-a,gennes1979a-a} and subsequent confirmation by
numerical methods, including careful series analysis of exact
enumerations \cite{guttmann1981a-a,berretti1985a-a} and statistical
analysis of high precision Monte Carlo simulations
\cite{grassberger1994a-a}, indicate that the upper critical dimension
for SAW is $d_u=4$. Above this dimension it is expected that SAW
behaviour is dominated by the same behaviour as occurs in Markovian
random walks (in a renormalisation group analysis of the O(0)
$\phi^4$ field theory both models are controlled by the so-called
Gaussian fixed point). We expect that while dominant exponents and
ratios of scaling amplitudes are the same as for RW the self-avoidance
constraint will affect scaling amplitudes and corrections to
scaling. In addition, apparently asymptotic $1/d$-expansions give
reasonable estimates of the connective constants. On the other hand,
much is known on a mathematically rigorous level
\cite{hara1992a-a,hara1992b-a,hara1993a-a,madras1993a-a}, thanks to the ingenuity of
the lace expansion, and so the similarity of pure random walk (RW) and
SAW behaviour can be quantified precisely to some extent.  Despite all
this non-rigorous and rigorous information several aspects of SAW in
high dimensions require numerical investigation. First, the connective
(or growth) constants, while bounded by rigorous arguments and
estimated by $1/d$-expansion values (and series analysis of fairly
short exact enumerations), are not known precisely from Monte Carlo
simulations, and so the relative value of different bounds is not well
understood. Second, the corrections to scaling in high dimensions have
not been investigated, and since controversies and intriguing findings
occur \cite{guim1997a-a} in low dimensions for the SAW and related
models, it is of interest to clarify these in high dimensions. In any
case, due to SAW being such a fundamental model it clearly is of
interest for us to establish as complete a description of the
behaviour of SAW as possible.

We have simulated self-avoiding walks on the $d$-dimensional
hypercubic lattice in dimensions $d=4$ to $d=8$ using the
Pruned-Enriched Rosenbluth Method (PERM), a clever generalisation of a
simple kinetic growth algorithm
\cite{grassberger1997a-a,frauenkron1998a-a} using a combination of
enrichment and pruning strategies to generate walks whose weights are
largely distributed around the expected peak of the distribution. We
have utilised an implementation similar to that described in
\cite{prellberg2000a-:a}, where the enrichment and pruning thresholds
are dynamically changed in response to the output of the algorithm
while maintaining a constant ratio between these thresholds.  For each
dimension $d=4,\ldots, 8$, we have generated $10^8$ samples of length
$N=4096$ and $10^7$ samples of length $N=16384$. While not having completely
independent samples we have crudely estimated the effect of the
dependence and so are able to give error estimates for our values. The
PERM algorithm is particularly appropriate for high-dimensional
simulations where SAW are close to RW, which are simply generated by a
Rosenbluth-Rosenbluth approach \cite{rosenbluth1955a-a}. While our
four-dimensional simulations build on the careful previous work of
Grassberger {\it et al.\ }\cite{grassberger1994a-a}, who simulated SAW
up to length 4000 on the four-dimensional hypercubic lattice, our
higher dimensional simulations are without parallel. In $d=4$, our simulations
do not provide any further insight
\cite{grassberger1994a-a,prellberg2000a-:a} into the subtle
logarithmic corrections predicted in four dimensions, but simply allow
us to update the connective and other constants in that dimension with
estimates that are an order of magnitude better than previously
obtained. In dimensions $d>4$, we compare our results to the bounds of
the lace expansion and other approaches, to the $1/d$-expansion, and to
series analysis of exact enumeration data. Our main results are
contained in Tables 1, 2 and 3. Apart from the precision of our
estimates, our other contribution is to point out evidence for
anomalous sub-dominant corrections to scaling in five dimensions
(which presumably occur in higher dimensions though so weakly as to be
not practically measurably).

Let the number of self-avoiding walks on the lattice of interest be
$c_N$, that is $c_N \equiv \left|\Omega_N\right|$ where $\Omega_N$ is
the set of all self-avoiding walks $\varphi$ of length $N$ steps
($N+1$ sites) with one end at some fixed origin. Let $p_N$ be the
number of self-avoiding polygons (closed walks) of length $N$. In this
paper we consider the $d$-dimensional hypercubic lattice for $d=4$, 5,
6, 7 and 8.  We define a reduced free energy or rather entropy per
step $\kappa_N$ as
\begin{equation}
\kappa_N={1\over N}\log c_N.
\end{equation}
Let $\langle Q \rangle_N$ denote the simple average of any quantity
$Q$ over the ensemble set of allowed paths $\Omega_N$ of length $N$.
Let $M(\varphi)$ be the number of non-consecutive nearest-neighbour
contacts (pairs of lattice sites occupied by the walk) for a given
walk $\varphi$.  We define a normalised mean number of contacts $m_N$
per step by
\begin{equation}
m_N = \frac{\langle M \rangle}{N}\, ,
\end{equation}
and a normalised fluctuation in the number of contacts per step by
\begin{equation}
f_N = \frac{\langle M^2 \rangle - \langle M \rangle^2}{N}\; .
\end{equation}
For a SAW model where each configuration is weighted by a Boltzmann
weight (say, $\omega^{M(\varphi)}$) to the number of nearest-neighbour contacts
(this model is known as interacting SAW or ISAW) the quantities $m_N$
and $f_N$ are proportional to the internal energy and specific heat in
the limit $\omega \rightarrow 1$. The thermodynamic limit for SAW is
given by the limit $N\rightarrow\infty$ so that the thermodynamic
limit entropy per step is given by
\begin{equation}
\kappa_\infty=\lim_{N \rightarrow \infty}\kappa_N \; .
\end{equation}
Given the thermodynamic limit exists this quantity determines the
partition function asymptotics, i.e. $c_N$ grows to leading order
exponentially as $\mu^N$ with the connective constant $\mu=e^{\kappa_\infty}$.

In our simulations we also calculated two measures of the walk's
average size.  Firstly, specifying a walk by the sequence of position
vectors ${{\bf r}_0, {\bf r}_1, ..., {\bf r}_N}$ the average
mean-square end-to-end distance is
\begin{equation}
\langle R^2_e\rangle_N = \langle ({\bf r}_N- {\bf r}_0) \cdot ({\bf r}_N- {\bf r}_0) \rangle\; .
\end{equation} 
We shall use the symbol $R^2_{e,N}$ to be equivalent to 
\begin{equation}
R^2_{e,N}\equiv \langle R^2_e\rangle_N .
\end{equation} 
The mean-square distance of the sites occupied by the walk from the
endpoint $\vecr_0$ of the walk is given by
\begin{equation}
\langle R_{m}^2 \rangle_N = \frac{1}{N+1}\sum_{i=0}^{N}\langle (\vecr_i - \vecr_0)\cdot(\vecr_i - \vecr_0)
\rangle\; .
\end{equation}
Again we define
\begin{equation}
R^2_{m,N}\equiv \langle R^2_m\rangle_N.
\end{equation}

Our main new results concern $d>4$, so we shall now discuss the
theoretical predictions for those dimensions.  It has been proved
\cite{hammersley1954a-a} that the thermodynamic limit exists, ie.  $\mu$
exists.  Furthermore it has been proved in sufficiently high
dimensions \cite{hara1992a-a,hara1992b-a} that
\begin{equation}
c_N = A \mu^N \left( 1 + O(n^{-\epsilon}) \right)
\end{equation}
for any  $\epsilon < \mathrm{min}\left( (d-4)/2,1 \right)$
and 
\begin{equation}
R^2_{e,N} = d_e \, N \, \left( 1 + O(n^{-\epsilon}) \right)
\end{equation}
for any  $\epsilon < \mathrm{min}\left( (d-4)/4,1 \right)$.

On the other hand on a non-rigorous level from the general theory of
critical phenomena \cite{li1995a-a} we further expect that the numbers
of walks and polygons have both analytic and non-analytic corrections
to scaling:
\begin{equation}
c_N  \sim  A \mu^N N^{\gamma -1} \left( 1 + {w_a\over N} +{w_e\over N^{\Delta_e}}  \right)
\end{equation}
and 
\begin{equation}
p_N  \sim  A \mu^N N^{\alpha - 2} \left( 1 + {p_a\over N} +{p_e\over N^{\Delta_e}}  \right)
\end{equation}
with $\gamma -1 =0$ and $\alpha - 2= -d/2$ for $d\geq 5$.
From this we deduce that the entropy, mean number of contacts and their
fluctuations scale as
\begin{equation}
\kappa_N \sim \kappa_\infty+ {k^{(1)}_a\over N} +{k^{(2)}_a\over N^2}+ {k_e\over N^{\Delta_e+1}}, 
\end{equation}
\begin{equation}
m_N \sim m_\infty+ {u^{(1)}_a\over N}+{u^{(2)}_a\over N^2} +{u_e\over N^{\Delta_e+1}} 
\end{equation}
and
\begin{equation}
f_N \sim f_\infty+ {s^{(1)}_a\over N}+{s^{(2)}_a\over N^2} +{s_e\over N^{\Delta_e+1}} 
\end{equation}
respectively. The non-analytic correction to scaling exponent
$\Delta_e$ is associated with the strongest non-analytic correction.
One would expect further non-analytic (other $\Delta$ exponent-like terms) and analytic
corrections (e.g.\ a $N^{-2}$ term)---see \cite{li1995a-a} for a more in-depth 
discussion of possible scaling forms in general dimensions. Our
numerical studies can discern only the strongest corrections to
scaling. 
Again for the geometric size of the walk one may hypothesise that
\begin{equation}
R^2_{e,N} \sim d_e\,N^{2\nu}\, \left(1+{e_a\over N}+{e_r\over N^{\Delta_r}} \right) 
\end{equation}
\begin{equation}
R^2_{m,N} \sim d_m\,N^{2\nu}\, \left(1+{o_a\over N}+{o_r\over N^{\Delta_r}} \right)
\end{equation}
with $2\nu=1$ for $d\geq 5$.

It may be tempting in a non-rigorous treatment of SAW above the upper
critical dimension to implicitly assume that the non-analytic
corrections to scaling either do not occur or only occur with the same
exponents as occur in RW. However, the field theoretic description of
critical phenomena above the upper critical dimension is subtle
(partially due to the presence of dangerously irrelevant variables)
and mean-field theory is unlikely to be the whole story. For example,
it is often assumed that hyperscaling relations break down above the
upper critical dimension. On the other hand, it is widely accepted
that the relation $2 - \alpha = d\nu$ holds for self-avoiding polygons
in all dimensions, where $2 - \alpha$ is the entropic exponent
associated $p_N$ and $\nu$ is the size exponent, in seeming
contradiction. It is certainly true that dominant exponents are
usually not controlled by the fluctuation dominated critical behaviour
that give rise to hyperscaling ($2 - \alpha \neq d\nu$ for the Ising
model for $d>4$) but rather mean-field energy vs entropy
physics. However, it may be that remnants of the fluctuation driven
critical behaviour still occur in high dimensions albeit now
contributing to the corrections to scaling.  In this picture the upper
critical dimension $d_u$ is the dimension below which fluctuation
driven critical phenomena (characterised by hyperscaling relations)
are dominant, while above $d_u$ they are sub-dominant to mean-field
criticality (fixed exponents).  One hyperscaling relation expected to
break down for $d \geq 5$ in SAW is $2\Delta_4 -\gamma = d\nu$ with
the ``gap'' exponent $\Delta_4$ associated with the ``intersection''
probability (see
\cite{li1995a-a,madras1993a-a} for example). One may be tempted to
hypothesise a correction to the scaling to $c_N$ of a term $\mu^N
N^{2\Delta_4 -d\nu-1}$. Since it is accepted that $\Delta_4=3/2$ (and
proved that $\nu=1/2$) we can hypothesise a correction to scaling
exponent arising from such a term as
\begin{equation}
\Delta_e={(d-4)\over 2}.
\end{equation}
That is, in $d=5$ we expect that $\Delta_e=1/2$, so corrections of
order $N^{-3/2}$ as well as analytic corrections of order $N^{-1}$,
$N^{-2}$ etc may occur. Unfortunately we are only practically able to
detect corrections of order $N^{-1}$ in our simulations and we have
been unable to detect even the $N^{-3/2}$ in $d=5$. We note in passing
that there are predicted logarithmic corrections in $d=4$ which we can
see, to a similar extent as in \cite{grassberger1994a-a}, in our
estimations of $\mu$---in fact we have utilised this expected
behaviour to give our refined estimation.  We now comment that the
non-analytic correction to scaling exponent hypothesised above is the
same as the ``crossover'' exponent $\phi_e=(d-4)/2$ of the Edwards
model. This leads us to conjecture that while there is strictly no
crossover from Gaussian to non-Gaussian behaviour, the excluded volume
could still make itself apparent in scaling through a scaling function
in the variable $bN^{\phi_e}$, where $b$ is the bare measure of the
excluded volume. We concede that a further assumption about the
expansion of the scaling function is needed here. In any case
following this line of argument it is then likely that such a
correction to scaling term occurs in other quantities such as in the
scaling of the size measures.

Assuming that the Edwards model crossover exponent provides the
dominant corrections to scaling exponent for the size
measures, $R^2_{e,N}$ and $R^2_{m,N}$, also gives us
\begin{equation}
\Delta_r=\Delta_e={(d-4)\over 2}.
\end{equation}
Hence in $d=5$ we expect that $\Delta_r=1/2$, and whenever the value of
$\Delta_r$ coincides with an analytic correction to scaling (e.g.\ $d=6$)
there may also be confluent logarithms appearing. We have been able to
successfully test for $\Delta_r=1/2$ in $d=5$ (see below), and we
even have some evidence of confluent logarithms present for $d=6$.
One can also predict that for large $N$ the quotients
\begin{equation}
B_N = R^2_{m,N}/R^2_{e,N} \sim B_\infty \left( 1+{b_a\over
N}+{b_r\over N^{\Delta_r}} \right) 
\end{equation}
approach the random walk value, ${B_\infty}= d_m/d_e = {1\over2}$ with
the same type of corrections as the size measures approach their
limits in $d\geq 5$.  In conclusion, from the scaling forms above we
predict that in $d=5$ we generically expect to see, within the quality
of data obtained, a correction of the order $N^{-1/2}$ in the size
measure quantities. In dimensions $7$ and $8$ we expect to see only
the $N^{-1}$ corrections, while in $d=6$ we may expect to see some
confluent logarithmic correction term such as $\log(N)/N$.

Our simulations allow us to estimate $\mu$, $m_\infty$, $f_\infty$,
$d_e$ (we certainly confirm that $d_e=2d_m$ in all dimensions studied)
and study the corrections to scaling in $\kappa_N$, $m_N$,
$R^2_{e,N}$, $R^2_{m,N}$, $B_N$. Let us first discuss the constants as
they provide the most important information contained in this
paper. Our best estimates and various comparisons are provided in
Tables \ref{tab1}, \ref{tab2} and \ref{tab3}. We are able to compare
our results to estimates and bounds from various sources. Fisher and
Gaunt \cite{fisher1964b-a} used exact enumerations on general
$d$-dimensional hypercubic lattices to give an asymptotic 
$1/d$-expansion for $\mu$ up to fifth order in the variable
$s=(2d-1)$. Nemirovsky and Freed \cite{nemirovsky1992a-a} extended
this to include $d_e$, while Ishinabe {\it et al.\
}\cite{ishinabe1994a-a} extended it to include $m_\infty$ and
$f_\infty$. In all these expansions the error is uncontrolled, and
since they are considered asymptotic expansions the optimal number of
terms to be used to give an accurate answer varies with dimension. All
terms have been used when applied to $d\geq 5$ since Fisher and Gaunt
\cite{fisher1964b-a} proposed that $d$ terms plus half the next
should be used in general. The $1/d$-expansions for $\mu$,
$m_\infty$, $f_\infty$, $d_e$, and $d_m$ are
\begin{eqnarray}
\mu&=&s-1/s-2/s^2-11/s^3-62/s^4+\ldots \\
m_\infty&=&1/s+1/s^2+7/s^3+35/s^4+250/s^5+\ldots\\
f_\infty&=&1/s+4/s^2+29/s^3+152/s^4+752/s^5+\ldots\\
d_e&=&1+2/s+28/s^2+180/s^3+1382/s^4+\ldots
\end{eqnarray}
where $s=(2d-1)$ and the expansion for $d_m$ is trivially given by
half of the expansion for $d_e$. The specific values for $d=4, \ldots,
8$ are given in Tables 1, 2 and 3. On the other hand, there has been
much effort expended to obtain rigorous and (semi-)rigorous upper and
lower bounds for the connective constants
\cite{hara1993a-a,slade1995a-a,alm1993a-a,noonan1998a-a,ponitz2000a-a} 
in all dimensions. The lace expansion provides excellent lower bounds
\cite{hara1993a-a,slade1995a-a} not only in high dimensions such as 5, 6, 7 and 8
but also in dimensions 3 and 4: lower bounds as quoted from Hara and
Slade \cite{hara1993a-a} for dimensions 4, 5 and 6, and for dimensions
7 and 8 have been computed via equation (2.34) of \cite{hara1993a-a},
are given in Table 1. The best current upper bounds
\cite{ponitz2000a-a,noonan1998a-a} are also listed in Tables 1, 2 and
3---the upper bounds for dimensions 7 and 8 have been computed using
the Maple code available at \cite{noonan1997a-a}.  We
also include in our tables the previous most precise estimates. 
In dimensions 5, 6, 7 and 8 these have been from
series analysis of exact enumeration data
\cite{guttmann1981a-a,nemirovsky1992b-a,douglas1995a-a}, while
in dimension 4 the previous best estimate of $\mu$ was obtained by Grassberger
\cite{grassberger1994a-a}. The previous estimates for $m_\infty$ are 
from exact enumerations and were given by Douglas and Ishinabe
\cite{douglas1995a-a}.

We now turn to our evidence for the types of corrections to scaling.
For the entropy, mean number of contacts, and fluctuations in the mean number
of contacts, we find that extrapolations assuming dominant $1/N$ corrections produce
consistent extrapolates in all dimensions $d\geq5$ and all lengths $N\geq 128$. 
Similarly for the size measure data, $R^2_{e,N}$, $R^2_{m,N}$ and $B_N$ in dimensions
$d=7$ and $d=8$ the assumption of $1/N$ corrections produces
consistent extrapolates for all lengths $N\geq128$. Thus we conclude 
that $1/N$ corrections dominate in those dimensions as predicted above. 

\begin{figure}[p]
\begin{center}
\hspace{-4.0cm}
\setlength{\unitlength}{0.240900pt}
\ifx\plotpoint\undefined\newsavebox{\plotpoint}\fi
\sbox{\plotpoint}{\rule[-0.200pt]{0.400pt}{0.400pt}}%
\begin{picture}(1050,900)(0,0)
\font\gnuplot=cmr10 at 10pt
\gnuplot
\sbox{\plotpoint}{\rule[-0.200pt]{0.400pt}{0.400pt}}%
\put(140.0,123.0){\rule[-0.200pt]{4.818pt}{0.400pt}}
\put(120,123){\makebox(0,0)[r]{1.41}}
\put(969.0,123.0){\rule[-0.200pt]{4.818pt}{0.400pt}}
\put(140.0,228.0){\rule[-0.200pt]{4.818pt}{0.400pt}}
\put(120,228){\makebox(0,0)[r]{1.42}}
\put(969.0,228.0){\rule[-0.200pt]{4.818pt}{0.400pt}}
\put(140.0,334.0){\rule[-0.200pt]{4.818pt}{0.400pt}}
\put(120,334){\makebox(0,0)[r]{1.43}}
\put(969.0,334.0){\rule[-0.200pt]{4.818pt}{0.400pt}}
\put(140.0,439.0){\rule[-0.200pt]{4.818pt}{0.400pt}}
\put(120,439){\makebox(0,0)[r]{1.44}}
\put(969.0,439.0){\rule[-0.200pt]{4.818pt}{0.400pt}}
\put(140.0,544.0){\rule[-0.200pt]{4.818pt}{0.400pt}}
\put(120,544){\makebox(0,0)[r]{1.45}}
\put(969.0,544.0){\rule[-0.200pt]{4.818pt}{0.400pt}}
\put(140.0,649.0){\rule[-0.200pt]{4.818pt}{0.400pt}}
\put(120,649){\makebox(0,0)[r]{1.46}}
\put(969.0,649.0){\rule[-0.200pt]{4.818pt}{0.400pt}}
\put(140.0,755.0){\rule[-0.200pt]{4.818pt}{0.400pt}}
\put(120,755){\makebox(0,0)[r]{1.47}}
\put(969.0,755.0){\rule[-0.200pt]{4.818pt}{0.400pt}}
\put(140.0,860.0){\rule[-0.200pt]{4.818pt}{0.400pt}}
\put(120,860){\makebox(0,0)[r]{1.48}}
\put(969.0,860.0){\rule[-0.200pt]{4.818pt}{0.400pt}}
\put(140.0,123.0){\rule[-0.200pt]{0.400pt}{4.818pt}}
\put(140,82){\makebox(0,0){0.00}}
\put(140.0,840.0){\rule[-0.200pt]{0.400pt}{4.818pt}}
\put(565.0,123.0){\rule[-0.200pt]{0.400pt}{4.818pt}}
\put(565,82){\makebox(0,0){0.05}}
\put(565.0,840.0){\rule[-0.200pt]{0.400pt}{4.818pt}}
\put(989.0,123.0){\rule[-0.200pt]{0.400pt}{4.818pt}}
\put(989,82){\makebox(0,0){0.10}}
\put(989.0,840.0){\rule[-0.200pt]{0.400pt}{4.818pt}}
\put(140.0,123.0){\rule[-0.200pt]{204.524pt}{0.400pt}}
\put(989.0,123.0){\rule[-0.200pt]{0.400pt}{177.543pt}}
\put(140.0,860.0){\rule[-0.200pt]{204.524pt}{0.400pt}}
\put(564,21){\makebox(0,0){$N^{-1/2}$}}
\put(565,702){\makebox(0,0)[l]{$R_{e,N}^2/N$}}
\put(140.0,123.0){\rule[-0.200pt]{0.400pt}{177.543pt}}
\put(890.0,164.0){\rule[-0.200pt]{0.400pt}{1.445pt}}
\put(880.0,164.0){\rule[-0.200pt]{4.818pt}{0.400pt}}
\put(880.0,170.0){\rule[-0.200pt]{4.818pt}{0.400pt}}
\put(771.0,265.0){\rule[-0.200pt]{0.400pt}{1.445pt}}
\put(761.0,265.0){\rule[-0.200pt]{4.818pt}{0.400pt}}
\put(761.0,271.0){\rule[-0.200pt]{4.818pt}{0.400pt}}
\put(671.0,349.0){\rule[-0.200pt]{0.400pt}{1.445pt}}
\put(661.0,349.0){\rule[-0.200pt]{4.818pt}{0.400pt}}
\put(661.0,355.0){\rule[-0.200pt]{4.818pt}{0.400pt}}
\put(586.0,423.0){\rule[-0.200pt]{0.400pt}{1.445pt}}
\put(576.0,423.0){\rule[-0.200pt]{4.818pt}{0.400pt}}
\put(576.0,429.0){\rule[-0.200pt]{4.818pt}{0.400pt}}
\put(515.0,485.0){\rule[-0.200pt]{0.400pt}{1.686pt}}
\put(505.0,485.0){\rule[-0.200pt]{4.818pt}{0.400pt}}
\put(505.0,492.0){\rule[-0.200pt]{4.818pt}{0.400pt}}
\put(456.0,539.0){\rule[-0.200pt]{0.400pt}{1.445pt}}
\put(446.0,539.0){\rule[-0.200pt]{4.818pt}{0.400pt}}
\put(446.0,545.0){\rule[-0.200pt]{4.818pt}{0.400pt}}
\put(405.0,585.0){\rule[-0.200pt]{0.400pt}{1.445pt}}
\put(395.0,585.0){\rule[-0.200pt]{4.818pt}{0.400pt}}
\put(395.0,591.0){\rule[-0.200pt]{4.818pt}{0.400pt}}
\put(363.0,624.0){\rule[-0.200pt]{0.400pt}{1.445pt}}
\put(353.0,624.0){\rule[-0.200pt]{4.818pt}{0.400pt}}
\put(353.0,630.0){\rule[-0.200pt]{4.818pt}{0.400pt}}
\put(328.0,658.0){\rule[-0.200pt]{0.400pt}{1.445pt}}
\put(318.0,658.0){\rule[-0.200pt]{4.818pt}{0.400pt}}
\put(318.0,664.0){\rule[-0.200pt]{4.818pt}{0.400pt}}
\put(298.0,684.0){\rule[-0.200pt]{0.400pt}{1.445pt}}
\put(288.0,684.0){\rule[-0.200pt]{4.818pt}{0.400pt}}
\put(288.0,690.0){\rule[-0.200pt]{4.818pt}{0.400pt}}
\put(273.0,706.0){\rule[-0.200pt]{0.400pt}{1.445pt}}
\put(263.0,706.0){\rule[-0.200pt]{4.818pt}{0.400pt}}
\put(890,167){\rule{1pt}{1pt}}
\put(771,268){\rule{1pt}{1pt}}
\put(671,352){\rule{1pt}{1pt}}
\put(586,426){\rule{1pt}{1pt}}
\put(515,489){\rule{1pt}{1pt}}
\put(456,542){\rule{1pt}{1pt}}
\put(405,588){\rule{1pt}{1pt}}
\put(363,627){\rule{1pt}{1pt}}
\put(328,661){\rule{1pt}{1pt}}
\put(298,687){\rule{1pt}{1pt}}
\put(273,709){\rule{1pt}{1pt}}
\put(263.0,712.0){\rule[-0.200pt]{4.818pt}{0.400pt}}
\end{picture}
\hspace{-0.5cm}
\setlength{\unitlength}{0.240900pt}
\ifx\plotpoint\undefined\newsavebox{\plotpoint}\fi
\begin{picture}(1050,900)(0,0)
\font\gnuplot=cmr10 at 10pt
\gnuplot
\sbox{\plotpoint}{\rule[-0.200pt]{0.400pt}{0.400pt}}%
\put(160.0,123.0){\rule[-0.200pt]{4.818pt}{0.400pt}}
\put(140,123){\makebox(0,0)[r]{0.496}}
\put(969.0,123.0){\rule[-0.200pt]{4.818pt}{0.400pt}}
\put(160.0,270.0){\rule[-0.200pt]{4.818pt}{0.400pt}}
\put(140,270){\makebox(0,0)[r]{0.497}}
\put(969.0,270.0){\rule[-0.200pt]{4.818pt}{0.400pt}}
\put(160.0,418.0){\rule[-0.200pt]{4.818pt}{0.400pt}}
\put(140,418){\makebox(0,0)[r]{0.498}}
\put(969.0,418.0){\rule[-0.200pt]{4.818pt}{0.400pt}}
\put(160.0,565.0){\rule[-0.200pt]{4.818pt}{0.400pt}}
\put(140,565){\makebox(0,0)[r]{0.499}}
\put(969.0,565.0){\rule[-0.200pt]{4.818pt}{0.400pt}}
\put(160.0,713.0){\rule[-0.200pt]{4.818pt}{0.400pt}}
\put(140,713){\makebox(0,0)[r]{0.500}}
\put(969.0,713.0){\rule[-0.200pt]{4.818pt}{0.400pt}}
\put(160.0,860.0){\rule[-0.200pt]{4.818pt}{0.400pt}}
\put(140,860){\makebox(0,0)[r]{0.501}}
\put(969.0,860.0){\rule[-0.200pt]{4.818pt}{0.400pt}}
\put(160.0,123.0){\rule[-0.200pt]{0.400pt}{4.818pt}}
\put(160,82){\makebox(0,0){0.00}}
\put(160.0,840.0){\rule[-0.200pt]{0.400pt}{4.818pt}}
\put(575.0,123.0){\rule[-0.200pt]{0.400pt}{4.818pt}}
\put(575,82){\makebox(0,0){0.05}}
\put(575.0,840.0){\rule[-0.200pt]{0.400pt}{4.818pt}}
\put(989.0,123.0){\rule[-0.200pt]{0.400pt}{4.818pt}}
\put(989,82){\makebox(0,0){0.10}}
\put(989.0,840.0){\rule[-0.200pt]{0.400pt}{4.818pt}}
\put(160.0,123.0){\rule[-0.200pt]{199.706pt}{0.400pt}}
\put(989.0,123.0){\rule[-0.200pt]{0.400pt}{177.543pt}}
\put(160.0,860.0){\rule[-0.200pt]{199.706pt}{0.400pt}}
\put(574,21){\makebox(0,0){$N^{-1/2}$}}
\put(575,713){\makebox(0,0)[l]{$R_{e,N}^2/R_{e,N}^2$}}
\put(160.0,123.0){\rule[-0.200pt]{0.400pt}{177.543pt}}
\put(893.0,178.0){\rule[-0.200pt]{0.400pt}{9.395pt}}
\put(883.0,178.0){\rule[-0.200pt]{4.818pt}{0.400pt}}
\put(883.0,217.0){\rule[-0.200pt]{4.818pt}{0.400pt}}
\put(776.0,271.0){\rule[-0.200pt]{0.400pt}{9.395pt}}
\put(766.0,271.0){\rule[-0.200pt]{4.818pt}{0.400pt}}
\put(766.0,310.0){\rule[-0.200pt]{4.818pt}{0.400pt}}
\put(678.0,355.0){\rule[-0.200pt]{0.400pt}{9.154pt}}
\put(668.0,355.0){\rule[-0.200pt]{4.818pt}{0.400pt}}
\put(668.0,393.0){\rule[-0.200pt]{4.818pt}{0.400pt}}
\put(596.0,411.0){\rule[-0.200pt]{0.400pt}{9.395pt}}
\put(586.0,411.0){\rule[-0.200pt]{4.818pt}{0.400pt}}
\put(586.0,450.0){\rule[-0.200pt]{4.818pt}{0.400pt}}
\put(526.0,458.0){\rule[-0.200pt]{0.400pt}{9.395pt}}
\put(516.0,458.0){\rule[-0.200pt]{4.818pt}{0.400pt}}
\put(516.0,497.0){\rule[-0.200pt]{4.818pt}{0.400pt}}
\put(468.0,498.0){\rule[-0.200pt]{0.400pt}{9.395pt}}
\put(458.0,498.0){\rule[-0.200pt]{4.818pt}{0.400pt}}
\put(458.0,537.0){\rule[-0.200pt]{4.818pt}{0.400pt}}
\put(419.0,531.0){\rule[-0.200pt]{0.400pt}{9.395pt}}
\put(409.0,531.0){\rule[-0.200pt]{4.818pt}{0.400pt}}
\put(409.0,570.0){\rule[-0.200pt]{4.818pt}{0.400pt}}
\put(378.0,563.0){\rule[-0.200pt]{0.400pt}{9.395pt}}
\put(368.0,563.0){\rule[-0.200pt]{4.818pt}{0.400pt}}
\put(368.0,602.0){\rule[-0.200pt]{4.818pt}{0.400pt}}
\put(343.0,587.0){\rule[-0.200pt]{0.400pt}{9.395pt}}
\put(333.0,587.0){\rule[-0.200pt]{4.818pt}{0.400pt}}
\put(333.0,626.0){\rule[-0.200pt]{4.818pt}{0.400pt}}
\put(314.0,611.0){\rule[-0.200pt]{0.400pt}{9.154pt}}
\put(304.0,611.0){\rule[-0.200pt]{4.818pt}{0.400pt}}
\put(304.0,649.0){\rule[-0.200pt]{4.818pt}{0.400pt}}
\put(290.0,621.0){\rule[-0.200pt]{0.400pt}{9.395pt}}
\put(280.0,621.0){\rule[-0.200pt]{4.818pt}{0.400pt}}
\put(893,198){\rule{1pt}{1pt}}
\put(776,290){\rule{1pt}{1pt}}
\put(678,374){\rule{1pt}{1pt}}
\put(596,430){\rule{1pt}{1pt}}
\put(526,477){\rule{1pt}{1pt}}
\put(468,517){\rule{1pt}{1pt}}
\put(419,551){\rule{1pt}{1pt}}
\put(378,583){\rule{1pt}{1pt}}
\put(343,607){\rule{1pt}{1pt}}
\put(314,630){\rule{1pt}{1pt}}
\put(290,641){\rule{1pt}{1pt}}
\put(280.0,660.0){\rule[-0.200pt]{4.818pt}{0.400pt}}
\end{picture}
\hspace{-4.0cm}
\caption{\it $R_{e,N}^2/N$ and $B_N=R_{m,N}^2/R_{e,N}^2$ versus $N^{-1/2}$ for $d=5$, showing clearly the presence
of the $N^{-1/2}$-correction to scaling. From the right-hand-side plot we extrapolate $B_\infty=0.5001(6)$.}
\label{fig1} 
\end{center}
\end{figure}
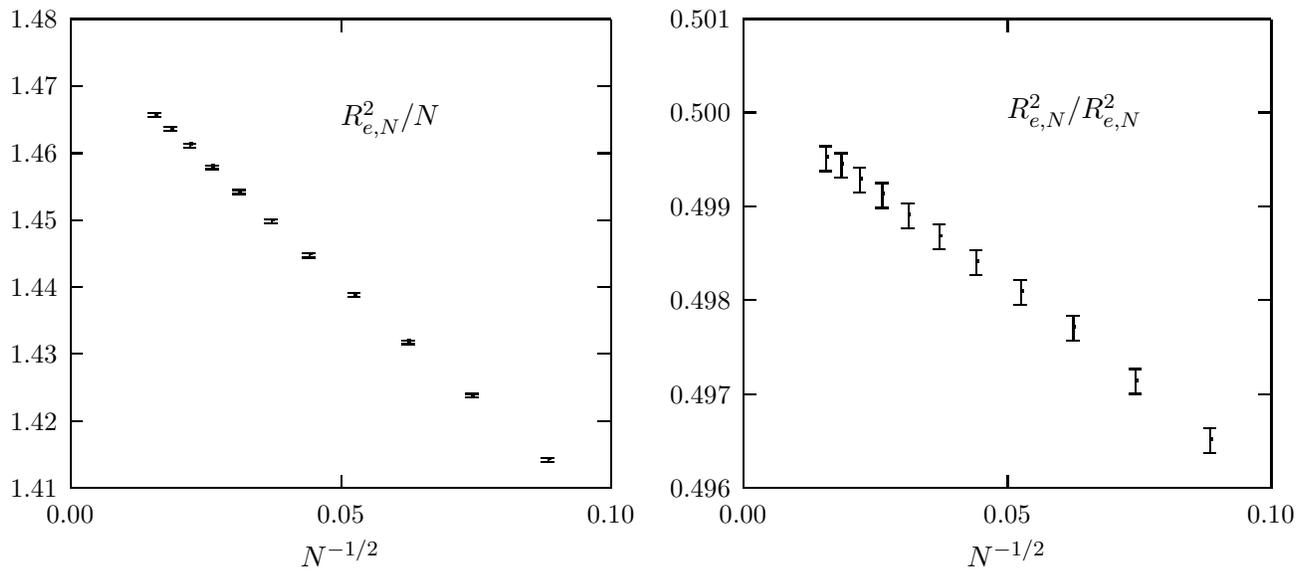

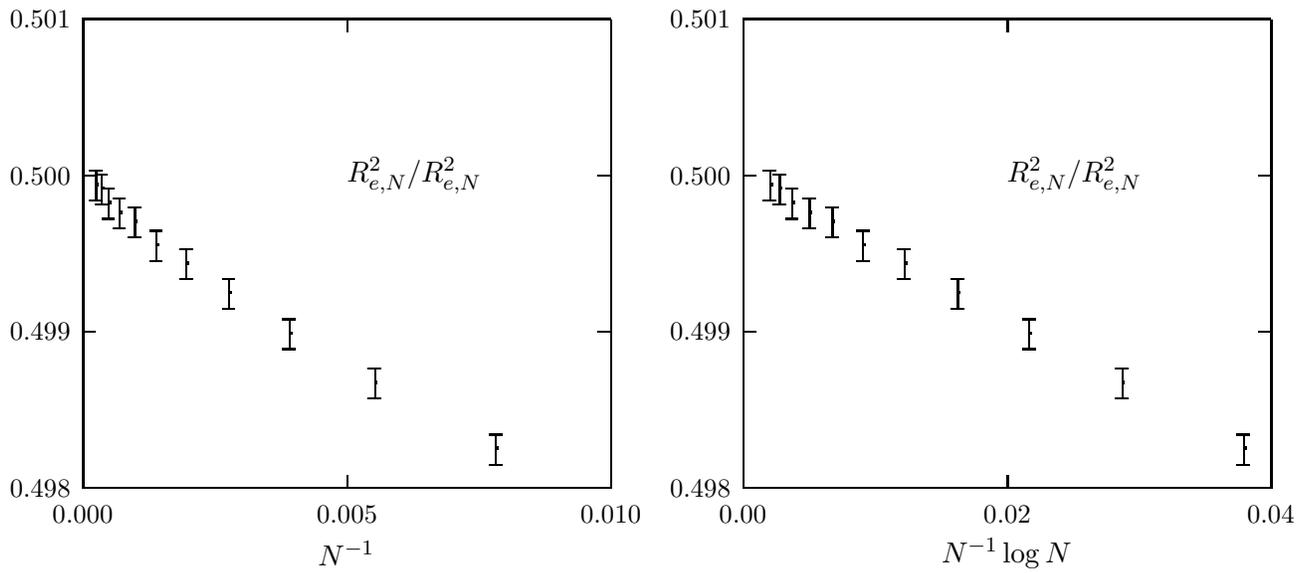
\begin{figure}[!p]
\begin{center}
\hspace{-4.0cm}
\setlength{\unitlength}{0.240900pt}
\ifx\plotpoint\undefined\newsavebox{\plotpoint}\fi
\sbox{\plotpoint}{\rule[-0.200pt]{0.400pt}{0.400pt}}%
\begin{picture}(1050,900)(0,0)
\font\gnuplot=cmr10 at 10pt
\gnuplot
\sbox{\plotpoint}{\rule[-0.200pt]{0.400pt}{0.400pt}}%
\put(160.0,123.0){\rule[-0.200pt]{4.818pt}{0.400pt}}
\put(140,123){\makebox(0,0)[r]{0.498}}
\put(969.0,123.0){\rule[-0.200pt]{4.818pt}{0.400pt}}
\put(160.0,369.0){\rule[-0.200pt]{4.818pt}{0.400pt}}
\put(140,369){\makebox(0,0)[r]{0.499}}
\put(969.0,369.0){\rule[-0.200pt]{4.818pt}{0.400pt}}
\put(160.0,614.0){\rule[-0.200pt]{4.818pt}{0.400pt}}
\put(140,614){\makebox(0,0)[r]{0.500}}
\put(969.0,614.0){\rule[-0.200pt]{4.818pt}{0.400pt}}
\put(160.0,860.0){\rule[-0.200pt]{4.818pt}{0.400pt}}
\put(140,860){\makebox(0,0)[r]{0.501}}
\put(969.0,860.0){\rule[-0.200pt]{4.818pt}{0.400pt}}
\put(160.0,123.0){\rule[-0.200pt]{0.400pt}{4.818pt}}
\put(160,82){\makebox(0,0){0.000}}
\put(160.0,840.0){\rule[-0.200pt]{0.400pt}{4.818pt}}
\put(575.0,123.0){\rule[-0.200pt]{0.400pt}{4.818pt}}
\put(575,82){\makebox(0,0){0.005}}
\put(575.0,840.0){\rule[-0.200pt]{0.400pt}{4.818pt}}
\put(989.0,123.0){\rule[-0.200pt]{0.400pt}{4.818pt}}
\put(989,82){\makebox(0,0){0.010}}
\put(989.0,840.0){\rule[-0.200pt]{0.400pt}{4.818pt}}
\put(160.0,123.0){\rule[-0.200pt]{199.706pt}{0.400pt}}
\put(989.0,123.0){\rule[-0.200pt]{0.400pt}{177.543pt}}
\put(160.0,860.0){\rule[-0.200pt]{199.706pt}{0.400pt}}
\put(574,21){\makebox(0,0){$N^{-1}$}}
\put(575,614){\makebox(0,0)[l]{$R_{e,N}^2/R_{e,N}^2$}}
\put(160.0,123.0){\rule[-0.200pt]{0.400pt}{177.543pt}}
\put(808.0,159.0){\rule[-0.200pt]{0.400pt}{11.563pt}}
\put(798.0,159.0){\rule[-0.200pt]{4.818pt}{0.400pt}}
\put(798.0,207.0){\rule[-0.200pt]{4.818pt}{0.400pt}}
\put(618.0,264.0){\rule[-0.200pt]{0.400pt}{11.322pt}}
\put(608.0,264.0){\rule[-0.200pt]{4.818pt}{0.400pt}}
\put(608.0,311.0){\rule[-0.200pt]{4.818pt}{0.400pt}}
\put(484.0,341.0){\rule[-0.200pt]{0.400pt}{11.322pt}}
\put(474.0,341.0){\rule[-0.200pt]{4.818pt}{0.400pt}}
\put(474.0,388.0){\rule[-0.200pt]{4.818pt}{0.400pt}}
\put(389.0,404.0){\rule[-0.200pt]{0.400pt}{11.322pt}}
\put(379.0,404.0){\rule[-0.200pt]{4.818pt}{0.400pt}}
\put(379.0,451.0){\rule[-0.200pt]{4.818pt}{0.400pt}}
\put(322.0,451.0){\rule[-0.200pt]{0.400pt}{11.322pt}}
\put(312.0,451.0){\rule[-0.200pt]{4.818pt}{0.400pt}}
\put(312.0,498.0){\rule[-0.200pt]{4.818pt}{0.400pt}}
\put(275.0,479.0){\rule[-0.200pt]{0.400pt}{11.563pt}}
\put(265.0,479.0){\rule[-0.200pt]{4.818pt}{0.400pt}}
\put(265.0,527.0){\rule[-0.200pt]{4.818pt}{0.400pt}}
\put(241.0,517.0){\rule[-0.200pt]{0.400pt}{11.322pt}}
\put(231.0,517.0){\rule[-0.200pt]{4.818pt}{0.400pt}}
\put(231.0,564.0){\rule[-0.200pt]{4.818pt}{0.400pt}}
\put(217.0,531.0){\rule[-0.200pt]{0.400pt}{11.322pt}}
\put(207.0,531.0){\rule[-0.200pt]{4.818pt}{0.400pt}}
\put(207.0,578.0){\rule[-0.200pt]{4.818pt}{0.400pt}}
\put(200.0,546.0){\rule[-0.200pt]{0.400pt}{11.322pt}}
\put(190.0,546.0){\rule[-0.200pt]{4.818pt}{0.400pt}}
\put(190.0,593.0){\rule[-0.200pt]{4.818pt}{0.400pt}}
\put(189.0,568.0){\rule[-0.200pt]{0.400pt}{11.322pt}}
\put(179.0,568.0){\rule[-0.200pt]{4.818pt}{0.400pt}}
\put(179.0,615.0){\rule[-0.200pt]{4.818pt}{0.400pt}}
\put(180.0,575.0){\rule[-0.200pt]{0.400pt}{11.322pt}}
\put(170.0,575.0){\rule[-0.200pt]{4.818pt}{0.400pt}}
\put(808,183){\rule{1pt}{1pt}}
\put(618,287){\rule{1pt}{1pt}}
\put(484,364){\rule{1pt}{1pt}}
\put(389,428){\rule{1pt}{1pt}}
\put(322,474){\rule{1pt}{1pt}}
\put(275,503){\rule{1pt}{1pt}}
\put(241,540){\rule{1pt}{1pt}}
\put(217,554){\rule{1pt}{1pt}}
\put(200,570){\rule{1pt}{1pt}}
\put(189,592){\rule{1pt}{1pt}}
\put(180,598){\rule{1pt}{1pt}}
\put(170.0,622.0){\rule[-0.200pt]{4.818pt}{0.400pt}}
\end{picture}
\hspace{-0.5cm}
\setlength{\unitlength}{0.240900pt}
\ifx\plotpoint\undefined\newsavebox{\plotpoint}\fi
\begin{picture}(1050,900)(0,0)
\font\gnuplot=cmr10 at 10pt
\gnuplot
\sbox{\plotpoint}{\rule[-0.200pt]{0.400pt}{0.400pt}}%
\put(160.0,123.0){\rule[-0.200pt]{4.818pt}{0.400pt}}
\put(140,123){\makebox(0,0)[r]{0.498}}
\put(969.0,123.0){\rule[-0.200pt]{4.818pt}{0.400pt}}
\put(160.0,369.0){\rule[-0.200pt]{4.818pt}{0.400pt}}
\put(140,369){\makebox(0,0)[r]{0.499}}
\put(969.0,369.0){\rule[-0.200pt]{4.818pt}{0.400pt}}
\put(160.0,614.0){\rule[-0.200pt]{4.818pt}{0.400pt}}
\put(140,614){\makebox(0,0)[r]{0.500}}
\put(969.0,614.0){\rule[-0.200pt]{4.818pt}{0.400pt}}
\put(160.0,860.0){\rule[-0.200pt]{4.818pt}{0.400pt}}
\put(140,860){\makebox(0,0)[r]{0.501}}
\put(969.0,860.0){\rule[-0.200pt]{4.818pt}{0.400pt}}
\put(160.0,123.0){\rule[-0.200pt]{0.400pt}{4.818pt}}
\put(160,82){\makebox(0,0){0.00}}
\put(160.0,840.0){\rule[-0.200pt]{0.400pt}{4.818pt}}
\put(575.0,123.0){\rule[-0.200pt]{0.400pt}{4.818pt}}
\put(575,82){\makebox(0,0){0.02}}
\put(575.0,840.0){\rule[-0.200pt]{0.400pt}{4.818pt}}
\put(989.0,123.0){\rule[-0.200pt]{0.400pt}{4.818pt}}
\put(989,82){\makebox(0,0){0.04}}
\put(989.0,840.0){\rule[-0.200pt]{0.400pt}{4.818pt}}
\put(160.0,123.0){\rule[-0.200pt]{199.706pt}{0.400pt}}
\put(989.0,123.0){\rule[-0.200pt]{0.400pt}{177.543pt}}
\put(160.0,860.0){\rule[-0.200pt]{199.706pt}{0.400pt}}
\put(574,21){\makebox(0,0){$N^{-1}\log N$}}
\put(575,614){\makebox(0,0)[l]{$R_{e,N}^2/R_{e,N}^2$}}
\put(160.0,123.0){\rule[-0.200pt]{0.400pt}{177.543pt}}
\put(946.0,159.0){\rule[-0.200pt]{0.400pt}{11.563pt}}
\put(936.0,159.0){\rule[-0.200pt]{4.818pt}{0.400pt}}
\put(936.0,207.0){\rule[-0.200pt]{4.818pt}{0.400pt}}
\put(755.0,264.0){\rule[-0.200pt]{0.400pt}{11.322pt}}
\put(745.0,264.0){\rule[-0.200pt]{4.818pt}{0.400pt}}
\put(745.0,311.0){\rule[-0.200pt]{4.818pt}{0.400pt}}
\put(609.0,341.0){\rule[-0.200pt]{0.400pt}{11.322pt}}
\put(599.0,341.0){\rule[-0.200pt]{4.818pt}{0.400pt}}
\put(599.0,388.0){\rule[-0.200pt]{4.818pt}{0.400pt}}
\put(497.0,404.0){\rule[-0.200pt]{0.400pt}{11.322pt}}
\put(487.0,404.0){\rule[-0.200pt]{4.818pt}{0.400pt}}
\put(487.0,451.0){\rule[-0.200pt]{4.818pt}{0.400pt}}
\put(413.0,451.0){\rule[-0.200pt]{0.400pt}{11.322pt}}
\put(403.0,451.0){\rule[-0.200pt]{4.818pt}{0.400pt}}
\put(403.0,498.0){\rule[-0.200pt]{4.818pt}{0.400pt}}
\put(348.0,479.0){\rule[-0.200pt]{0.400pt}{11.563pt}}
\put(338.0,479.0){\rule[-0.200pt]{4.818pt}{0.400pt}}
\put(338.0,527.0){\rule[-0.200pt]{4.818pt}{0.400pt}}
\put(300.0,517.0){\rule[-0.200pt]{0.400pt}{11.322pt}}
\put(290.0,517.0){\rule[-0.200pt]{4.818pt}{0.400pt}}
\put(290.0,564.0){\rule[-0.200pt]{4.818pt}{0.400pt}}
\put(264.0,531.0){\rule[-0.200pt]{0.400pt}{11.322pt}}
\put(254.0,531.0){\rule[-0.200pt]{4.818pt}{0.400pt}}
\put(254.0,578.0){\rule[-0.200pt]{4.818pt}{0.400pt}}
\put(237.0,546.0){\rule[-0.200pt]{0.400pt}{11.322pt}}
\put(227.0,546.0){\rule[-0.200pt]{4.818pt}{0.400pt}}
\put(227.0,593.0){\rule[-0.200pt]{4.818pt}{0.400pt}}
\put(217.0,568.0){\rule[-0.200pt]{0.400pt}{11.322pt}}
\put(207.0,568.0){\rule[-0.200pt]{4.818pt}{0.400pt}}
\put(207.0,615.0){\rule[-0.200pt]{4.818pt}{0.400pt}}
\put(202.0,575.0){\rule[-0.200pt]{0.400pt}{11.322pt}}
\put(192.0,575.0){\rule[-0.200pt]{4.818pt}{0.400pt}}
\put(946,183){\rule{1pt}{1pt}}
\put(755,287){\rule{1pt}{1pt}}
\put(609,364){\rule{1pt}{1pt}}
\put(497,428){\rule{1pt}{1pt}}
\put(413,474){\rule{1pt}{1pt}}
\put(348,503){\rule{1pt}{1pt}}
\put(300,540){\rule{1pt}{1pt}}
\put(264,554){\rule{1pt}{1pt}}
\put(237,570){\rule{1pt}{1pt}}
\put(217,592){\rule{1pt}{1pt}}
\put(202,598){\rule{1pt}{1pt}}
\put(192.0,622.0){\rule[-0.200pt]{4.818pt}{0.400pt}}
\end{picture}
\hspace{-4.0cm}
\caption{\it $B_N=R_{m,N}^2/R_{e,N}^2$ versus $N^{-1}$ and versus $N^{-1}\log N$ for $d=6$, showing the possible
presence of a confluent logarithm for the $N^{-1}$-correction to scaling. We extrapolate $B_\infty=0.5000(2)$.}
\label{fig2} 
\end{center}
\end{figure}

The evidence for anomalous scaling is summarised in Figures \ref{fig1}
and \ref{fig2}: for $d=5$, we clearly detect $N^{-1/2}$-corrections
(see Figure
\ref{fig1}) in the size measure data, and for $d=6$ our results are suggestive of 
$N^{-1}\log N$-corrections, which produce a slightly better fit than
$N^{-1}$-corrections (see Figure \ref{fig2}). We note in passing that the absence of
$N^{-1/2}$-corrections in past extrapolations from exact enumeration
data (see Table \ref{tab3}) may have affected previous estimates for
$A_e$.

In summary, we have presented a comprehensive study of scaling of
self-avoiding walks at and above the upper critical dimension, testing
various scaling predictions and providing precise estimates of
associated scaling amplitudes.

\section*{Acknowledgements} 
Financial support from the Australian Research Council is gratefully
acknowledged by ALO. We thank Richard Brak for his comments on the
manuscript.


\newpage

\begin{table}
\begin{center}
\begin{tabular}{l|lllll}
\hline
dimension& 4& 5& 6& 7& 8\\
\hline
\hline
estimate for $\mu$ & 6.774043(5)& 8.838544(3)& 10.878094(4)& 12.902817(3)& 14.919257(2)\\
\hline
previous estimates&  6.77404(4)&  8.8386(8)& 10.8788(9)& 12.900&14.920\\
\cite{grassberger1994a-a,guttmann1981a-a,nemirovsky1992b-a}&&&&\\
lower bound\cite{hara1993a-a}&  6.7429& 8.8285& 10.8740& 12.8811 & 14.9030\\
upper bound\cite{ponitz2000a-a,noonan1998a-a}&  6.8040 &  8.8602& 10.8886& 12.9081& 14.9221\\
$1/d$-expansion\cite{fisher1964b-a}&  6.7714& 8.8397&10.8800&12.9040& 14.9200\\
\hline
\end{tabular}
\caption{\it Numerical values for self-avoiding walk connective constants $\mu$ in dimension 4 to 8. 
}
\label{tab1} 
\end{center}
\end{table}

\clearpage
\newpage

\begin{table}
\begin{center}
\begin{tabular}{l|lllll}
\hline
dimension& 4& 5& 6& 7& 8\\
\hline
\hline
estimate for $m_\infty$& 0.17088(5)& 0.134576(6)&0.106902(4) &0.087715(2) & 0.074222(2)\\
\hline
previous estimate\cite{douglas1995a-a}&  0.1740(15)&  0.141(1)&  0.111(1)&  0.0892(8)& 0.0744(6)\\
$1/d$-expansion\cite{douglas1995a-a}& 0.213125 & 0.142627 & 0.108376 & 0.087925& 0.074206\\
\hline
estimate for $f_\infty$& 0.330(2)& 0.2324(3)& 0.1640(2)&0.12331(7) &0.09818(5) \\
\hline
$1/d$-expansion\cite{douglas1995a-a}& 0.417 & 0.2362 & 0.1608 & 0.12114& 0.09703\\
\hline
\end{tabular}
\caption{\it Numerical values for normalised mean $m_\infty$ and fluctuation $f_\infty$
 of nearest-neighbour contacts in dimension 4 to 8.}
\label{tab2} 
\end{center}
\end{table}

\clearpage
\newpage

\begin{table}
\begin{center}
\begin{tabular}{l|lllll}
\hline
dimension& 4& 5& 6& 7& 8\\
\hline
\hline
estimate for $d_e$&---& 1.4767(13)& 1.2940(6)& 1.2187(3)& 1.1760(2)\\
\hline
previous estimate\cite{nemirovsky1992a-a}&---&  1.434&  1.296&  1.222&  1.178\\
$1/d$-expansion\cite{nemirovsky1992a-a}&---&  1.385&  1.273&  1.212&  1.174\\
\hline
estimate for $d_m$& --- & 0.7385(6)& 0.6470(2)& 0.6094(1)& 0.5880(1)\\
estimate for $B_\infty=d_m/d_e$&0.504(7)& 0.5001(6)& 0.5000(2)& 0.5000(1)& 0.5000(1)\\
\hline
\end{tabular}
\caption{\it Numerical values for distance amplitudes $d_m$ and $d_e$ and their quotient
$B_\infty=d_m/d_e$ in dimension 4 to 8.}
\label{tab3} 
\end{center}
\end{table}

\clearpage
\newpage

\clearpage
\newpage

\end{document}